\documentclass[a4paper,11pt]{article}
\pdfoutput=1

\usepackage{jheppub}

\usepackage{amsmath,amssymb,amsfonts}
\usepackage{graphicx}
\usepackage{xcolor}
\usepackage{mathrsfs}
\usepackage{slashed}
\usepackage[utf8]{inputenc}
\usepackage[english]{babel}
\usepackage{subfigure}
\usepackage{mdframed}
\usepackage{ulem}
\usepackage{units}

\hypersetup{colorlinks,citecolor=blue,urlcolor=blue,linkcolor=blue}

\def\beq{\begin{equation}\begin{aligned}}
\def\eeq{\end{aligned}\end{equation}}

\begin{document}

\title{An Intermediate Scale R-axion \& the QCD Axion}

\author{James Unwin}
\affiliation{Department of Physics, University of Illinois Chicago, Chicago, IL 60607, USA}

\abstract{
An intermediate scale R-axion faces an immediate obstruction from the Dine-Festuccia-Komargodski (DFK) bound on the superpotential, $2|\langle W\rangle|\leq f_R F$, since for a nearly Minkowski vacuum it typically follows that $f_R\gtrsim M_{\rm Pl}$.  We show that this lower bound on $f_R$ can be relaxed in an effective construction with the scalar potential tuned near zero via a mixed $F$- and $D$-term uplift,  leading to a metastable vacuum in which the usual Planckian-$f_R$ inference from the DFK argument is avoided locally.
Validity of the effective field theory and metastability of the small $f_R$ vacuum generically both imply a relaxed bound: $ f_R  \gtrsim \sqrt{m_{3/2}M_{\rm Pl}} $.  We also highlight that if the R-symmetry has a QCD anomaly, this potentially permits the R-axion to play the role of the QCD axion. TeV-scale supersymmetry permits $f_R\sim10^{11}$ GeV, this not only evades certain astrophysical and cosmological axion constraints, but notably lies in the window for which the observed dark matter abundance can be reproduced by the R-axion via the misalignment mechanism.}

\setcounter{tocdepth}{1}

\maketitle

\section{Introduction}

It is a generic expectation that a topological term should appear in the QCD Lagrangian of the form \cite{Crewther:1979pi}
\beq
 {\mathcal L}_\theta
 =
 \frac{\alpha_s}{8\pi}
 \bar\theta 
 G^a_{\mu\nu}\widetilde G^{a\mu\nu}.
\eeq
This term is dressed by the effective QCD vacuum angle $ \bar\theta =  \theta+\arg\det M_q$, where $\theta$ is the bare QCD angle and $M_q$ is the quark mass matrix. It is unreasonable to expect these two contributions to cancel to high precision.  One would instead expect $|\bar\theta|={\mathcal O}(1)$, in contrast, the non-observation of a neutron electric dipole moment \cite{Pendlebury:2015lrz} requires $|\bar\theta|\lesssim10^{-10}$. This is the strong CP problem. 

The Peccei-Quinn (PQ) mechanism solves the strong CP problem by promoting $\bar\theta$ to a dynamical field \cite{Peccei:1977hh,Peccei:1977ur}.  A spontaneously broken anomalous global U(1) gives rise to a pseudo-Nambu-Goldstone (pNGB) boson, the QCD axion, whose potential is generated by QCD and whose minimum dynamically sets $\bar\theta$ to zero.  In invisible axion models, such as DFSZ and KSVZ \cite{Zhitnitsky:1980tq,Dine:1981rt,Kim:1979if,Shifman:1979if}, this U(1) symmetry is imposed on new scalar or fermionic fields.  While this successfully resolves the strong CP problem, in many cases, the PQ symmetry appears in a somewhat \textit{ad hoc} manner.  It is thus interesting to ask whether a symmetry already present in motivated Standard Model extensions can play the role of the PQ symmetry.

One natural candidate in supersymmetric (SUSY) theories is a continuous U(1)$_R$ symmetry.  An R-symmetry is distinct from ordinary internal symmetries, as it acts on the superspace coordinates $ \vartheta\rightarrow e^{i\alpha}\vartheta$, and thus rotates the supercharges.  Accordingly, for the superpotential to be invariant in the action, it is required to carry R-charge two, with scalar and fermion components of a chiral multiplet having different R-charges.  

Notably, if U(1)$_R$ is anomalous under QCD and spontaneously broken, it is a candidate for the PQ mechanism. In this case the R-axion is identified with the QCD axion. This is attractive because the new states carrying U(1)$_R$ are simply the superpartners. The QCD anomaly arises naturally since the gluino and any additional R-charged colored matter fields will contribute to the mixed U(1)$_R$-SU(3)$_c^2$ anomaly, and is non-zero for appropriate field content and $R$-charge assignments.

However, there are two major obstructions to realizing the PQ mechanism via U(1)$_R$.  The first obstacle is that in conventional supergravity theories, the cosmological constant is commonly tuned by adding a constant term $W_0$ to the superpotential.  Since the superpotential carries R-charge two, $W_0\neq0$  explicitly breaks the R-symmetry.  This explicit R-breaking is much larger than that due to the QCD anomaly, and as a result, the PQ mechanism is spoiled, shifting the axion minimum away from the CP-conserving point. In a recent paper  \cite{Unwin:2024yqq} we highlighted a special example in which the cosmological constant is tuned to zero in an R-symmetric manner via a non-minimal K\"ahler potential \cite{Claudson:1983cr}. This model provided a \textit{proof of principle} that this issue can be evaded. 

A drawback of the R-symmetric tuning of \cite{Unwin:2024yqq,Claudson:1983cr} is that the R-breaking scale is found to be at the Planck scale. Indeed, Planck scale R-breaking can be understood within a more general statement, which we highlight as the second major obstruction. Specifically, Dine-Festuccia-Komargodski (DFK) \cite{Dine:2009sw} demonstrated that for spontaneous $F$-term SUSY breaking and a spontaneously broken continuous R-symmetry, the superpotential expectation value is bounded. Moreover, if one further assumes a nearly Minkowski vacuum, this leads to a strong bound on the R-breaking scale:
\beq
 f_R\gtrsim M_{\rm Pl} .
\label{eq:0}
\eeq
Planck-scale R-breaking leads to a light R-axion which can be cosmologically problematic. Further, in the case that the QCD anomaly is the dominant source of explicit breaking, the R-axion follows the QCD axion mass scaling, with $f_R\sim M_{\rm Pl} $ implying $m_a\sim 10^{-12}$~eV. Such ultralight axions not only lead to problematic cosmological overproduction (which can potentially be circumvented in special constructions \cite{Dvali:1995ce,Dvali:2026ceb}), but a QCD axion in this range is also constrained by superradiance bounds from the non-observation of black hole spin-down \cite{Arvanitaki:2009fg,Ning:2026ebu}. 
The aim of this paper is to evade the Planckian bound of eq.~(\ref{eq:0}) within an effective field theory (EFT), and thus obtain an intermediate scale  $f_R$ that is not in conflict with cosmological and astrophysical limits. Notably, for an intermediate $f_R\sim 10^{11}$ GeV, the R-axion becomes a viable dark matter candidate.

The paper is structured as follows: In Section~\ref{S2} we discuss the DFK bound and its implications for the R-axion, and construct a mixed $F/D$ SUSY-breaking model that permits an intermediate scale $f_R$. In Section~\ref{S3}, we examine the vacuum conditions and their implications for the uplift sector. Metastability of the vacuum is examined in Section~\ref{S5}, from which we derive a new lower bound on $f_R$. Section~\ref{S6} discusses the prospect of identifying the R-axion with the QCD axion within this special construction. 
The limitation of our construction is that, being an effective theory, it requires UV completion. Moreover, realizing some of the necessary UV conditions may be non-trivial, as we discuss within the concluding remarks presented in Section~\ref{S7}.

\section{Bounds on the R-axion decay constant}
\label{S2}

The DFK bound on the superpotential \cite{Dine:2009sw} begins from the fact that, for a continuous U(1)$_R$ symmetry, the superpotential $W$ has R-charge two.  In an ordinary four-dimensional description with chiral fields $\Phi^i$ transforming linearly under the R-symmetry, this implies
\beq
   R_i\Phi^i\partial_iW = 2W ,
 \label{eq:1}
\eeq
where $R_i$ is the R-charge of $\Phi^i$.
It follows (even for non-canonical K\"ahler potentials) from the Cauchy-Schwarz inequality that  \cite{Dine:2009sw}
\beq
 2|\langle W\rangle| \leq  f_R F ,
  \label{eq:DFK}
\eeq
where $f_R$ is the R-axion decay constant and $F$ is the SUSY-breaking $F$-term. 
Setting the vacuum to be nearly Minkowski requires the $F$ term to be parametrically $F\sim   m_{3/2}M_{\rm Pl}$, in which case the gravitino mass is set by 
\beq
 m_{3/2}
 =
 e^{K/(2M_{\rm Pl}^2)}
 \frac{|\langle W\rangle|}{M_{\rm Pl}^2}.
\eeq
Substituting these into eq.~\eqref{eq:DFK}, yields eq.~(\ref{eq:0}), namely  the R-breaking scale is required to be at the Planck scale:
$ f_R\gtrsim M_{\rm Pl} $. The reason the DFK bound is so restrictive in ordinary supergravity is that the SUSY-breaking $F$-term is also responsible for canceling the vacuum-energy contribution from $|\langle W\rangle|$.  Since one requires $F\sim m_{3/2}M_{\rm Pl}$ to have a nearly Minkowski vacuum, the DFK inequality forces the R-breaking to be Planckian.

Notably, in the supergravity limit the covariant form of the DFK inequality involves the projection of the K\"ahler-covariant derivative along the R-Goldstone direction
\beq
 |X_R^iD_iW|
 \leq
 f_R F ,
   \label{eq:DFK2}
\eeq
where $X_R^i$ is the holomorphic Killing vector generating the R-transformation $\delta\Phi^i=\alpha X_R^i$.
For a linearly transforming chiral field of R-charge ($R_i$), one has
\beq
X_R^i=iR_i\Phi^i .
\eeq
Since the superpotential has R-charge two $W(\Phi)\rightarrow e^{2i\alpha}W(\Phi)$, infinitesimally $\delta W=2i\alpha W$. Moreover, from an application of the chain rule one also has
\beq
\delta W=\partial_iW\delta\Phi^i=\alpha X_R^i\partial_iW .
\eeq
Equating these two forms of $\delta W$ we obtain
\beq
X_R^i\partial_iW=2iW .
\eeq
Then examining  eq.~(\ref{eq:DFK2}) and noting that in the rigid SUSY limit $D_iW\rightarrow\partial_iW$, it follows
\beq
|X_R^iD_iW|
\rightarrow
|X_R^i\partial_iW|
=
2|W| ,
\eeq
which recovers the left-hand side (LHS) of the usual DFK inequality, given in eq.~(\ref{eq:DFK}).

In the construction of Section~\ref{S3}, the only field shifted by the R-symmetry is a modulus field $T$,
hence
\beq
X_R^iD_iW = X_R^TD_TW = iD_TW .
\eeq
We identify a special metastable minimum for which $D_TW=0$, and as a result the covariant DFK inequality reduces locally to
\beq
 0\leq f_R F .
\label{0}
\eeq
Consequently, this no longer leads to the Planckian lower bound on $f_R$.

Thus, whilst the DFK bound of eq.~(\ref{eq:DFK}) is seemingly very robust, the subsequent implication $f_R\gtrsim M_{\rm Pl}$ can potentially be evaded. The core idea is that, unlike the $ F$-term, a $D$-term is not tied to the R variation of the superpotential (cf.~eq.~(\ref{eq:1})), thus the $D$-term can participate in the vacuum-energy cancellation without directly setting $f_R$. This approach is motivated by $D$-term only breaking constructions such as \cite{Dvali:1997sf}.
We find a viable model in a mixed $F/D$ term uplift that gives a metastable Minkowski vacuum, in which $F$-terms of the (field-dependent) uplift stabilize the saxion, but such that the $F$ term along the R-breaking direction vanishes at the minimum.\footnote{The simpler model with only a constant positive $D$-term fails, as we outline in Appendix \ref{SA}.  Specifically, the small $f_R$ point is a supersymmetric AdS extremum, but the saxion is unavoidably tachyonic. } We note upfront that we only provide an effective construction, and we do not fully specify the dynamics of the uplift sector.

We consider a modulus $T$ with shift symmetry $T\rightarrow T+i\alpha$, which transforms non-linearly under the R-symmetry such that its imaginary part becomes the R-axion. This type of nonlinear transform has previously been considered in R-symmetric modulus/axion models, see e.g.~\cite{Harigaya:2014ola} (also compare with \cite{Dine:2009sw,Antoniadis:2014hfa}). Additionally, we introduce a single chiral field $\phi$ with R-charge zero and charge $+1$ under a gauged U(1)$_X$.  The field $T$ is neutral under U(1)$_X$, so a U(1)$_X$ gauge transformation does not shift $\tau={\rm Im} T$ and the R-axion is not eaten.  We consider a superpotential of the form
\beq
 W=W_*e^{2T},
\label{W}
 \eeq
where  $W_*=\langle W \rangle$ is the scale of the superpotential expectation value, hence the scale of the gravitino mass and SUSY breaking after uplift.
We take $W_*$ to have R-charge zero, thus $W_*$ does not break the R-symmetry.  The full superpotential $W$ transforms appropriately (with R-charge 2) under U(1)${}_R$ because the modulus transforms non-linearly $T\rightarrow T+i\alpha$ leading to $W\rightarrow W_*e^{2(T+i\alpha)} = e^{2i\alpha}W$.

We consider a K\"ahler potential of the following form
\beq
 K =  K^{(0)} +  |\phi|^2 ,
 \quad {\rm with} \quad
 K^{(0)} = M_{\rm Pl}^2 \left[-2(T+T^\dagger) +\frac{\epsilon^2}{2}(T+T^\dagger)^2+\frac{c}{4}(T+T^\dagger)^4 \right].
\label{K}
\eeq
We parameterize $T=s+i\tau$ such that the saxion field is $s\equiv (T+T^\dagger)/2$. 
The kinetic term for $T=s+i\tau$ is therefore
\beq
 {\cal L}_{\rm kin}
 =  K_{T\bar T} \partial_\mu T \partial^\mu T^\dagger= K_{T\bar T}
 \left[(\partial_\mu s)^2+ (\partial_\mu\tau)^2\right].
 \label{kin}
\eeq
Since the R-transformation shifts $\tau\rightarrow\tau+\alpha$, the canonically normalized R-axion is 
\beq
 a_R=\sqrt{2K_{T\bar T}} ~\tau .
 \label{cann}
\eeq
Taking derivatives along the $T$ direction we have
\beq
   K_T  =M_{\rm Pl}^2 \left( -2  + 2\epsilon^2s +  8cs^3\right),
 \qquad
 K_{T\bar T} =M_{\rm Pl}^2 \left(  \epsilon^2+12cs^2\right) ,
\label{KTT}
\eeq
and, by comparison to eq.~(\ref{cann}), we identify
\beq
 f_R(s)= \sqrt{2K_{T\bar T}} =  \sqrt{2}M_{\rm Pl}
 \sqrt{ \epsilon^2+12cs^2} .
 \label{fR}
\eeq
Notably, at $s=0$ the R-axion decay constant reduces to
\beq
 f_R(0)
 =
 \sqrt{2}\epsilon M_{\rm Pl}.
\label{fR0}
\eeq
While for $\epsilon\sim1$ this leads to a Planckian $f_R$, for $\epsilon\ll1$ an intermediate scale R-axion can be realized. This small decay constant only holds at the special point $s=0$ and away from this point, assuming $c\sim1$ the decay constant is again $f_R(s)\sim M_{\rm Pl}$.  Notably, however, one cannot make $f_R$ arbitrarily small, first we show below that EFT validity restricts $f_R$ and then in Section~\ref{S5} we will demonstrate that metastability of the vacuum also constrains $f_R$.
 
That one cannot take $f_R$ arbitrarily small within the EFT can be seen from the requirement that the saxion remains well described by the EFT. The saxion-dependent axion kinetic term is given by eq.~(\ref{kin}) and depends on $K_{T\bar T}(s)$. Thus, in order to establish the local stability of the small-$f_R$ vacuum within the effective theory, the saxion cannot be integrated out from the start. Near $s=0$ the canonically normalized saxion and axion are 
\beq
 \sigma & = \sqrt{2K_{T\bar T}(0)} s =\sqrt{2}\epsilon M_{\rm Pl}s=f_R s,
 \\
 a_R &  = \sqrt{2K_{T\bar T}(0)}\,\tau= \sqrt{2}\epsilon M_{\rm Pl}\tau= f_R\tau .
\label{hat} \eeq
Once canonically normalized, the kinetic terms imply a cross-term (coming from $K_{T\bar T}(s)$) of the form
\beq\label{La}
 {\cal L}_{a}= \frac{1}{2}(\partial_\mu a_R)^2+ \frac{1}{2}(\partial_\mu \sigma)^2+\frac{1}{\Lambda^2} \sigma^2(\partial_\mu a_R)^2 +\frac{1}{\Lambda^2} \sigma^2(\partial_\mu \sigma)^2 ~,
\eeq
which we write in terms of an EFT cutoff scale $\Lambda^{-2}=12c (M_{\rm Pl}^2/f_R^4)$.

If the saxion mass exceeds the cutoff $\Lambda$, calculations of the vacuum stability within the EFT are no longer reliable. Thus, validity of the EFT  requires that  $m_\sigma<\Lambda$. Moreover, the saxion mass is parametrically the gravitino mass $m_{3/2}$ and hence we can express this EFT restriction as a characteristic bound on the R-axion decay constant
\beq
 f_R \gtrsim\sqrt{m_{3/2}M_{\rm Pl}}.
 \label{EFT}
\eeq
In the next section we show $m_\sigma^2\simeq m_{3/2}^2$  and identify the leading order dependence of the saxion mass, after which we refine the EFT bound above. Notably, this constraint on $f_R$  is much weaker than the typical bound coming from standard DFK arguments. Moreover, for a TeV scale gravitino, this EFT bound implies that $f_R$ can be intermediate scale. 

 Thus the construction is controlled by two quantities: $K_{T\bar T}$ sets the size of $f_R$, and $D_TW$ determines whether $T$ participates in SUSY breaking. In the case $D_TW=0$ the cosmological constant is set to zero via the uplift sector (rather than $T$) and the Minkowski condition no longer forces $f_R\sim M_{\rm Pl}$.
 To examine how this construction leads to $D_TW=0$ at vacuum, we consider $s=0$ as a candidate minimum and parametrize the uplift by the effective $D$-term potential
\beq
 V_D
 =
 \frac{g^2}{2}{\mathcal D}_X^2,
  \quad {\rm with} \quad
   {\mathcal D}_X
 =  |\phi|^2+\xi .
 \label{eq:D}
\eeq
This should be treated as an effective description. While the constant Fayet-Iliopoulos (FI) parameter $\xi$  \cite{Fayet:1974jb} is useful for parametrizing the uplift, genuinely constant FI terms are constrained in supergravity \cite{Freedman:1976uk,Barbieri:1982ac,Stelle:1978wj,Komargodski:2009pc,Cribiori:2017laj,Elvang:2006jk,Binetruy:2004hh}. The $D$-term is connected to how charged fields transform and to the K\"ahler-covariant derivatives of the superpotential via \cite{Komargodski:2009pc}
\beq
 {\mathcal D}_X
 \propto
 \frac{X_X^iD_iW}{W}.
 \label{eq:Killing}
\eeq
Notably, eq.~(\ref{eq:Killing}) implies that for $\langle W\rangle\neq0$, a nonzero $D$-term is generally accompanied by nonzero $F$-terms in the fields charged under U(1)$_X$ \cite{Dumitrescu:2010ca}. Thus, the form of eq.~(\ref{eq:D}) typically requires UV completion with the constant $\xi$ replaced by field-dependent $\xi(\Phi_i)$, a St\"uckelberg mechanism, or Green-Schwarz mechanism \cite{Dvali:1996rj,Arkani-Hamed:1998ufq} (see also \cite{Unwin:2024yqq}).

We next show explicitly that the uplift sector necessarily carries a nonzero $F$-term, even though $\phi$ does not appear in the superpotential.
Since $\phi$ is absent from $W$ (cf.~eq.~(\ref{W})) but appears in the K\"ahler potential, we have
\beq
  D_\phi W  =
 \partial_\phi W
 +   \frac{K_\phi}{M_{\rm Pl}^2}W
 =   \frac{\phi^\dagger}{M_{\rm Pl}^2}W ~,
 \label{eq:DphiW}
\eeq
where we use that $ K_\phi=\phi^\dagger$.
Thus $F_\phi\propto D_\phi W$ is nonzero at the uplifted minimum, where $\phi\neq0$.  The vacuum is therefore set by a mix of $F$- and $D$-term breaking.   
Despite this uplift-sector SUSY breaking, the $T$ sector itself remains supersymmetric at the candidate minimum.
This can be seen by direct calculation of $D_TW$ at the candidate minimum $s=0$.
From eq.~\eqref{KTT} we have that at $s=0$ 
\beq
 K_T\big|_{s=0}
 =  -2M_{\rm Pl}^2,
 \qquad
 \partial_TW\big|_{s=0}
 =   2W .
\eeq
Hence, the K\"ahler covariant derivative of $W$ vanishes at $s=0$
\beq
 D_TW
 = \partial_TW
 + \frac{K_T}{M_{\rm Pl}^2}W
 = 0 .
\label{2.9}\eeq
Furthermore, it follows that at $s=0$ the projection of the K\"ahler-covariant derivative along the R-Goldstone direction $X_R^iD_iW$ vanishes 
\beq
 X_R^iD_iW = iD_TW   =  0 .
 \label{main}
\eeq
For the nonlinear R-transformation $ T\rightarrow T+i\alpha$, an infinitesimal transform gives $\delta T = i\alpha$, thus $X_R^T=i $ (while $X_R^\phi=0$).  The uplift sector breaks SUSY, but the SUSY-breaking direction is orthogonal to the R-Goldstone direction, so the large vacuum energy canceling order parameter does not force the R-axion decay constant to be Planckian.

This condition $D_TW = 0$ is the central reason the usual DFK argument can be avoided locally, as outlined in eqns.~\eqref{eq:DFK2}-\eqref{0}. Notably, $D_TW=0$ is realized in our model at the special point which simultaneously sets $K_{T\bar T}=\epsilon^2M_{\rm Pl}^2$, permitting an intermediate scale $f_R$ for $\epsilon\ll1$. Thus, it remains to show that $s=0$ is indeed a (metastable) minimum.

 \section{Vacuum structure and tuning the cosmological constant}
\label{S3}

Having identified the necessary conditions for realizing an intermediate scale $f_R$, we next examine the vacuum structure and verify that $s=0$ is  a minimum of the scalar potential.
The scalar potential associated to $W$ and $K$ given by eq.~(\ref{W}) \& (\ref{K}) can be written as
\beq
V=V_F+V_D.
\eeq
The $F$-term contribution is given by
\beq \label{VF}
V_F
&=e^{K/M_{\rm Pl}^2}
\left( K^{T\bar T}D_TW D_{\bar T}\overline W
 + K^{\phi\bar\phi}D_\phi W D_{\bar\phi}\overline W - 3\frac{|W|^2}{M_{\rm Pl}^2} \right)\\
&= m_{3/2}^2 M_{\rm Pl}^2 \left[\frac{4s^2(\epsilon^2+4cs^2)^2}{\epsilon^2+12cs^2}+r-3\right],
\eeq
where we define 
\beq
 r\equiv\frac{|\phi|^2}{M_{\rm Pl}^2}.
 \eeq 
Away from vacuum the gravitino mass is a function $m_{3/2}=m_{3/2}(s,r)$ with
\beq
m_{3/2}^2(s,r)
=
\exp\!\left[
r+2\epsilon^2s^2+4cs^4
\right]
\frac{|W_*|^2}{M_{\rm Pl}^4}.
\eeq
 The $D$-term contribution (given by eq.~(\ref{eq:D})) can be expressed
\beq
V_D
=\frac{g^2}{2}
\left( r M_{\rm Pl}^2+\xi\right)^2 .
\eeq
At $s=0$ we have $K^{(0)}=0$, thus $K=|\phi|^2$ and the gravitino mass is given by
\beq
m_{3/2}^2
=
e^r
\frac{|W_*|^2}{M_{\rm Pl}^4}.
 \label{eq:m32}
\eeq
We use $m_{3/2}$ (without showing the explicit dependences) to indicate the vacuum gravitino mass.
Putting this together, the scalar potential at $s=0$ is
\beq
 V(r)|_{s=0}
 = \left( r-3 \right)e^r \frac{|W_*|^2}{M_{\rm Pl}^2} +  \frac{g^2}{2}  \left(rM_{\rm Pl}^2+\xi  \right)^2 .
\label{Vs0}
\eeq

We can now identify the uplifted Minkowski vacuum by imposing the stationarity condition (in $r$) and the vanishing of the vacuum energy. First, we identify the $r=r_\star$ such that $ V_{s=0}(r_\star)=0$, which  implies the condition 
\beq
 \frac{g^2}{2}
 \left(
  r_\star M_{\rm Pl}^2+\xi
 \right)^2
 =
 (3-r_\star)m_{3/2}^2M_{\rm Pl}^2 .
 \label{eq:Mink}
\eeq
The $D$-term uplift (LHS) must cancel against the difference between the positive $\phi$ $F$-term and the negative supergravity contribution. For this to be possible, it is required that 
\beq
 r_\star<3.
\label{rs3}
 \eeq
Furthermore, the stationarity condition is
\beq
 \left. \frac{\partial V_{s=0}}{\partial r}
\right|_{r=r_\star}
= (r_\star-2)e^{r_\star} \frac{|W_*|^2}{M_{\rm Pl}^2}
+ g^2M_{\rm Pl}^2 \left(  r_\star M_{\rm Pl}^2+\xi \right) =0 
\eeq
which implies
\beq
 g^2M_{\rm Pl}^2
 \left(
  r_\star M_{\rm Pl}^2+\xi
 \right)
 =
 -(r_\star-2)m_{3/2}^2M_{\rm Pl}^2 .
 \label{eq:stat}
\eeq
We use eq.~\eqref{eq:stat} to substitute for $(r_\star M_{\rm Pl}^2+\xi)$ in eq.~\eqref{eq:Mink} and solve for $g$ to obtain
\beq
g^2
=\left(\frac{m_{3/2}}{M_{\rm Pl}}\right)^2\frac{(r_\star-2)^2}{2(3-r_\star)} . 
\label{g}\eeq
Thus the magnitude of the U(1)$_X$ gauge coupling is determined by $r_\star$ and the SUSY breaking scale $m_{3/2}$.
From eq.~\eqref{eq:stat} we can also obtain the form of the FI parameter
\beq
 \xi  =   -r_\star M_{\rm Pl}^2  -  \frac{(r_\star-2)m_{3/2}^2}{g^2}.
\label{xi}\eeq
We note that in this effective description, $r_\star$ is not a derived quantity, rather, it labels the chosen stationary point of the uplift sector.  Once $r_\star$ is specified by the uplift sector, the Minkowski and stationarity conditions then determine $g$ and $\xi$. The value of $r_\star$ is bounded from above by $ r_\star<3$ in order for the $D$-term uplift to have the necessary sign to cancel against the supergravity and $\phi$ $F$-term contributions. 

In the above we assumed that a stable minimum could be found at $s=0$, we next verify this, paying particular attention to the saxion.  Notably, saxion stabilization is precisely where the pure $D$-term lift model fails, as outlined in the Appendix \ref{SA}.
  At fixed $r$, the $D$-term in
eq.~\eqref{eq:D} is independent of $s$, while the $F$-term potential is
\beq
 V_F(s) =m_{3/2}^2
 M_{\rm Pl}^2 \exp\left[ 2\epsilon^2s^2+4cs^4\right] 
 \left[ \frac{4s^2(\epsilon^2+4cs^2)^2}{\epsilon^2+12cs^2}+ r-3\right].
 \label{eq:VFx}
\eeq
Expanding the $F$-term potential near $s=0$ gives
\beq
 V_F(s) \simeq V_F(0)+2m_{3/2}^2M_{\rm Pl}^2(r-1)\epsilon^2s^2+\cdots .
\eeq
This coefficient is not yet the physical saxion mass, since $s={\rm Re} T$ is not canonically normalized. Recalling the canonically normalized saxion $ \sigma = \sqrt{2}\epsilon M_{\rm Pl}s $ (given in eq.~(\ref{hat}))  we rewrite the quadratic term in terms of $\sigma$ to obtain
\beq
 V_F \simeq V_F(0) + (r_\star-1)m_{3/2}^2\sigma^2+\cdots .
\eeq
Comparing with $V\simeq V(0)+\frac12m_\sigma^2\sigma^2$, one obtains
\beq
 m_\sigma^2= 2(r_\star-1)m_{3/2}^2 .
 \label{ms}
\eeq
The saxion is locally stabilized for $r>1$. For $r<1$ the saxion is tachyonic, similar to the constant $D$-term uplift. 
 Thus for the critical point at $s=0$ to be a (local) minimum it is required that the VEV of $\phi$ to lie in the range
\beq
 1<r_\star<3 ,
 \label{13}
\eeq
with the upper bound coming from eq.~(\ref{rs3}). Finally, in Appendix \ref{ApA} we examine the radial uplift direction and fully confirm that the uplifted stationary point is a local minimum in the range $1<r_\star<3$.

Combining the requirement that $r_\star>1$ with eq.~\eqref{g}, we observe that the U(1)${}_X$ gauge coupling must generically be very small $ g  \lesssim \frac{m_{3/2}}{M_{\rm Pl}}$. A representative example is given by $r_\star\simeq1.10$ with $|\langle\phi\rangle|\simeq1.05M_{\rm Pl}$, in which case the fraction of the positive vacuum energy contribution which the $D$-term is responsible for is $(3-r_\star)/3\simeq0.63$, with the remaining positive contribution coming from the $\phi$ $F$-term.
Additionally, it follows from eq.~(\ref{xi}) that generically $|\xi|\sim M_{\rm Pl}^2$.
Furthermore, using the full form of eq.~(\ref{ms}) in the EFT constraint of eq.~(\ref{EFT}) gives a more complete picture of the EFT validity requirement:
\beq
 f_R
 \gtrsim
 \left[
  \sqrt{24c(r_\star-1)}
  m_{3/2}M_{\rm Pl}
 \right]^{1/2},
\label{EFT2}
\eeq
with the factor of $\sqrt{12c}$ from the definition of $\Lambda$ in eq.~\eqref{La}.

We note two limitations of this effective construction; firstly, the field $\phi$ has a Planckian expectation value (similar to that encountered in \cite{Claudson:1983cr}). This implies that higher-order Planck-induced corrections in the K\"ahler potential for $\phi$ are unsuppressed and thus the vacuum is not under full control without specifying a UV completion. A second limitation is that the uplift typically requires an extremely weak gauge sector with $ g\sim m_{3/2}/M_{\rm Pl}$.
This is not inconsistent within the EFT, but it may be constrained in any quantum gravity completion, via weak-gravity-type considerations \cite{Arkani-Hamed:2006emk}. One caveat is that if $r_\star\approx3$ then $g$ can be large; specifically, writing $r_\star=3-\delta$, then the Minkowski condition implies
\beq
g^2=\frac{m_{3/2}^2}{2\delta M_{\rm Pl}^2} .
\label{eps}
\eeq
For $g\sim1$ with TeV SUSY this means that the uplift sector should stabilize $r_\star=3-\delta$ with $\delta\sim10^{-30}$. 
This is an additional tuning, separate from the tuning of the vacuum energy. There could conceivably be some dynamical mechanism that could push $r_\star$ near this special value, but we do not explore this possibility further here.
%


\section{Metastability bound on the decay constant}
\label{S5}

The effective construction above realizes a Minkowski vacuum with mixed $F$- and $D$-term SUSY breaking. As highlighted in eq.~(\ref{fR0}), for $\epsilon\ll1$ and $1<r_\star<3$, the model yields a parametrically small R-axion decay constant at $s=0$ with
\beq
 f_R(s)\big|_{s=0}=\sqrt{2}\epsilon M_{\rm Pl}\ll M_{\rm Pl}.
\eeq
However, this vacuum is only locally stable.  To study whether this is a global minimum, one must allow the uplift field to readjust as the saxion $s$ is
displaced from zero. Along this trajectory, $r$ varies to minimize the energy at each value of $s$. 

Specifically, we consider the trajectory in field space defined by
\beq
 V_{\rm eff}(s)
 =
 \min_r V(s,r).
\eeq
We argue in Appendix \ref{ApB} that generically\footnote{Note that the special case $r_\star\approx3$ is distinct, and in this scenario $s=0$ likely provides the global minimum, as we outline in Appendix \ref{ApD}.} there is a deeper region away from $s=0$, towards $s\sim\mathcal{O}(1)$ and along this trajectory the minimum evolves toward smaller values of $r$, which corresponds to smaller $|\phi|$. 

Notably, away from $s=0$ the cancellation that leads to $D_TW=0$ (cf.~eq.~(\ref{2.9})) no longer occurs. This can be seen by noting that while $\partial_TW   =   2W$ (for all $s$), the derivative of the K\"ahler potential differs with $s$ \beq
K_T\big|_{s\neq0}
\simeq
M_{\rm Pl}^2\left(-2+2s\epsilon^2+8cs^3\right).
\eeq
Explicitly evaluating $D_TW$ for $s\neq0$
\beq
 D_TW=\partial_TW+\frac{K_T}{M_{\rm Pl}^2}W
 = 2s(\epsilon^2+4cs^2)W ,
\eeq
confirms that for $s_0={\cal O}(1)$ one has $D_TW\neq0$. The non-vanishing of $D_TW$ means that the $T$ multiplet now participates in supersymmetry breaking and, as a result, the usual DFK implication for the R-axion reasserts itself, leading to $f_R\sim M_{\rm Pl}$.
We can see this explicitly by noting that the same displacement of $s$ from zero also makes $f_R$ Planckian 
\beq
 f_R(s) = \sqrt{2K_{T\bar T}}
 = \sqrt{2}M_{\rm Pl} \sqrt{\epsilon^2+12cs^2}~.
\eeq

Near the special point $s=0$ the uplift is from mixed sources, with both the $D$-term and the $\phi$ $F$-term contributing to the positive vacuum energy,  However, along the lower-energy trajectory at larger $s$, the $\phi$ contribution decreases.  The system therefore moves toward a regime that more closely resembles the (unsuccessful) constant-uplift construction discussed in Appendix~\ref{SA}. Although the small-$f_R$ Minkowski vacuum near $s=0$ is only a local minimum, it may still be metastable. We note in passing that long-lived metastable vacua are common in supersymmetric theories, for instance, in ISS type models \cite{Intriligator:2006dd}, and in mixed $F/D$ breaking models, e.g.~\cite{Dienes:2008gj}.  Insisting that the small-$f_R$  vacuum is sufficiently long-lived to be phenomenologically viable will constrain the model, as we discuss next.

To calculate the lifetime of the metastable vacuum at $s=0$ we  compute the Euclidean bounce action for tunneling from the small-$f_R$ false vacuum to the lower-energy Planckian-$f_R$ region at $s\sim1$ (following e.g.~\cite{Coleman:1977py,Callan:1977pt,Duncan:1992ai,Intriligator:2006dd,Adams:1993zs}).
The height of the potential barrier that separates the small-$f_R$ point from the more favourable Planckian-$f_R$ region is parametrically small.
  Expanding eq.~\eqref{eq:VFx} at fixed $r$ and small $s$ gives
\beq
V_F(s)-V_F(0)\simeq m_{3/2}^2M_{\rm Pl}^2\left[2(r-1)\epsilon^2s^2-4(7-r)cs^4+\frac{256c^2}{\epsilon^2}s^6+\cdots\right].
\label{expa}
\eeq
For the range of interest $1<r_\star<3$, the quadratic term is positive while the quartic term is negative, providing a  barrier separating the local minimum at $s=0$ from the deeper region at $s\sim1$.  Taking just these two terms, we approximate
\beq
 V_F(s)-V_F(0)
 \simeq
 m_{3/2}^2M_{\rm Pl}^2 \left(  A s^2- B s^4\right),
\eeq
with
$ A=2(r-1)\epsilon^2$ and $B=4(7-r)c $. The saxion displacement at the top of the barrier $s_{\rm b}$ can be found by solving
\beq
 \frac{d}{ds}  \left( A s^2-Bs^4  \right)= 0,
\eeq
which gives
\beq
 s_{\rm b}^2=\frac{A}{2B}=\frac{(r-1)\epsilon^2}{4(7-r)c}.
\eeq
We note that in the above, we have used just the first two terms of eq.~\eqref{expa}. Since the barrier occurs at $s^2\sim\epsilon^2$, the expansion parameter $cs^2/\epsilon^2$ is order one, and higher terms in the expansion are not parametrically suppressed. This should be sufficient to identify the $s_{\rm b}$ scaling, but it is unreliable for obtaining numerical prefactors.

Proceeding, we substitute $s_{\rm b}$ into the truncated potential to obtain
\beq
 \Delta V_{\rm barrier}^{(r)}= m_{3/2}^2M_{\rm Pl}^2  \left[ A s_{\rm b}^2 -B s_{\rm b}^4
  \right]  =m_{3/2}^2M_{\rm Pl}^2  \frac{A^2}{4B}.
\eeq
 Since $\epsilon$ controls the splitting between $f_R$ and $M_{\rm Pl}$, cf.~eq.~\eqref{fR}, and since we take $c\sim1$, the barrier is parametrically
\beq
 \Delta V_{\rm barrier}
 \sim
 m_{3/2}^2M_{\rm Pl}^2\epsilon^4
  \sim   m_{3/2}^2M_{\rm Pl}^2 \left(   \frac{f_R}{M_{\rm Pl}}   \right)^4 .
\eeq
The transition is in the thick-wall regime and the bounce estimate is set by \cite{Duncan:1992ai,Adams:1993zs,Intriligator:2006dd}
\beq  
S_4  \sim  \frac{(\Delta\sigma)^4}{\Delta V_{\rm barrier}},
\eeq
in terms of the displacement to the barrier of the canonically normalized saxion $\Delta\sigma$.  Recalling the canonical normalization of eq.~\eqref{hat}, one has
\beq
 \Delta\sigma \sim \epsilon M_{\rm Pl}s_{\rm b}  \sim \frac{\epsilon^2M_{\rm Pl}}{\sqrt c} \sim \frac{f_R^2}{M_{\rm Pl}\sqrt c}.
\eeq
Note that since the field excursion is sub-Planckian gravitational corrections are negligible \cite{Coleman:1980aw}.
It follows that the bounce action is parametrically
\beq
 S_4 \sim
 \epsilon^4\frac{M_{\rm Pl}^2}{m_{3/2}^2}   \sim  \left(  \frac{f_R^2}{M_{\rm Pl}m_{3/2}} \right)^2 .
\eeq

Requiring the false vacuum to be longer lived than the age of the Universe is commonly stated as the restriction: $ S_4\gtrsim400 $ \cite{Coleman:1977py,Callan:1977pt}. This can be translated into the following bound on the R-axion decay constant
\beq
 f_R
 \gtrsim 400^{1/4}
 \sqrt{m_{3/2}M_{\rm Pl}}
 \simeq
 {\rm few}\times 10^{11}~{\rm GeV}
 \left(
  \frac{m_{3/2}}{1~{\rm TeV}}
 \right)^{1/2}.
 \label{eq:fR-bound}
\eeq
Thus, TeV scale SUSY permits an intermediate scale $f_R$. The price of an intermediate scale $f_R$ is a metastable vacuum, a small parameter $\epsilon\ll1$, and certain special conditions must be met in the UV completion.


\section{A QCD R-axion}
\label{S6}

Before concluding, we discuss the possibility that an intermediate scale R-axion may play the role of the QCD axion.\footnote{We note the interpretation of the QCD R-axion construction in \cite{Unwin:2024yqq} should be refined. That model assumed the R-axion mass contribution from hidden sector spontaneous R-breaking could be sequestered, but since U(1)${}_R$ is a single cross-sector symmetry, every R-charged VEV contributes to the same R-Goldstone. The visible sector nevertheless contains an accidental PQ symmetry in addition to the full R-symmetry, so the spectrum contains both an ultralight R-axion, with $f_R\sim M_{\rm Pl}$, and a separate intermediate scale PQ axion, with $f_a\sim10^{11} {\rm GeV}$.  While this implies that the construction in \cite{Unwin:2024yqq}  does not realize the original goal of identifying a phenomenologically viable model with the R-axion identified with the QCD axion, it remains notable that the PQ symmetry arises accidentally from the global R-symmetry and thus may be protected from Planck-suppressed operators if the R-symmetry is gauged in the UV completion to supergravity.}
As outlined in the introduction, this is attractive because an anomalous U(1)$_R$ naturally arises in SUSY extensions of the Standard Model and automatically has the ingredients needed to realize an invisible QCD axion scenario. 
One of the major obstructions to using U(1)$_R$ within the PQ mechanism is the DFK bound that typically prohibits intermediate scale $f_R$. Since we have constructed a special scenario which allows for an intermediate scale R-axion we should re-examine the prospect of a QCD R-axion within this framework.
We highlight first that the mixed $F$- and $D$-term uplift described above sets the cosmological constant to zero without explicit R-violation. In this manner, it is similar to the Claudson-Hall-Hinchliffe R-symmetric model \cite{Claudson:1983cr}, the difference being that the model presented here allows $f_R$ to be parametrically below $M_{\rm Pl}$.

Since there is no explicit R-breaking the R-axion is a massless Goldstone boson of the spontaneously broken U(1)$_R$, prior to QCD effects. If the U(1)$_R$-SU(3)$_c^2$ anomaly is non-zero this explicitly violates U(1)$_R$ (as in the PQ mechanism) and is the leading source of the R-axion mass with $m_a\propto 1/f_a$ where $f_a$ is the physical QCD axion decay constant. 
The low-energy QCD coupling is also determined by the mixed U(1)$_R$-SU(3)$_c^2$ anomaly.  We define the anomaly coefficient $N_R$ by
\beq
 \partial_\mu J_R^\mu
 \supset
 \frac{\alpha_s}{8\pi}
 N_R G^a_{\mu\nu}\widetilde G^{a\mu\nu}.
\eeq

For binary R-charges $R(Q)=R(U^c)=R(D^c)=1$, and $R(H_u)=R(H_d)=0$, the MSSM anomaly coefficient is $N_R^{\rm binary}=3$. However, the Dirac partner $\Phi_3$ introduced for the gluino with R($\Phi_3$)=0 contributes  $\Delta N_R=-3$. The gluino Dirac partner $\Phi_3$ is necessary for phenomenologically viable TeV scale SUSY, but in this case  $N_R=
N_R^{\rm binary}+\Delta N_R=0$. This suggests two options \cite{Unwin:2024yqq}: 
\begin{itemize}
\item[i).]~High scale SUSY with the gluinos obtaining highly suppressed loop-induced Majorana masses after R-breaking and removing $\Phi_3$; 
\item[ii).]~TeV scale SUSY with non-binary R-charges for the MSSM fields, retaining $\Phi_3$.
\end{itemize}
Taking the most general R-charges for the MSSM fields compatible with Yukawa couplings and the Weinberg operator, the net anomaly including $\Phi_3$ can be calculated to be \cite{Unwin:2024yqq}
\beq
N_R=-\frac{3}{2}R(H_uH_d).
\label{HuHd}
\eeq
Thus U(1)$_R$ is anomalous provided the R-charge of the bilinear operator $H_uH_d$ is non-zero. 

\newpage

The low energy effective Lagrangian is given by
\beq
 {\mathcal L}
 \supset
 \frac{\alpha_s}{8\pi}
 \frac{N_Ra_R}{f_R}
 G^a_{\mu\nu}\widetilde G^{a\mu\nu}.
\eeq
The physical QCD axion decay constant is therefore
\beq
 f_a
 =
 \frac{f_R}{|N_R|}.
\eeq
The anomaly coefficient $N_R$  receives contributions from all colored fermions charged under U(1)$_R$, including the gluino. 
Provided $N_R\neq0$ the U(1)$_R$ will dynamically set $\bar\theta=0$ and the (QCD) R-axion will acquire a mass given by \cite{Srednicki:1985xd}
\beq
 m_a
 \simeq
 60 \mu{\rm eV}
 \left(
  \frac{10^{11} {\rm GeV}}{f_a}
 \right),
\eeq
where the reference value for $f_a$ is indicative of $f_a\simeq f_R$ and draws on the TeV SUSY lower bound of eq.~(\ref{eq:fR-bound}).  Moreover, it is interesting that $f_a\sim10^{11} {\rm GeV}$ lies comfortably within the window favoured for QCD axion dark matter produced via the misalignment mechanism, which is normally quoted as $10^9\lesssim f_a\lesssim 10^{12}$ GeV \cite{Preskill:1982cy}. We note that for post-inflationary PQ breaking, domain walls lead to cosmological issues. To avoid this domain wall problem, the anomaly coefficient should be set to $|N_R|=1$ (i.e.~$R(H_uH_d)=2/3$ for the simplest model of eq.~(\ref{HuHd})) or introduce additional dynamics to remove the wall network  \cite{Sikivie:1982qv}.

Aside from being an interesting symmetry, U(1)${}_R$ also elegantly resolves an internal hierarchy issue within the Minimally Supersymmetric Standard Model (MSSM), namely the $\mu$ problem. Specifically, if symmetries permit the term  $ \mu H_uH_d$ in the Lagrangian, the natural expectation is $\mu\sim M_{\rm Pl}$. With appropriate R-charge assignment the R-symmetry forbids the $\mu$-term, this term can be subsequently reintroduced at higher order in a manner that makes it naturally TeV scale. In particular, the $\mu$-term can arise after R-breaking through a (Kim-Nilles type \cite{Kim:1994eu}) higher dimensional operator
\beq
 W_{\rm KN}
 =\kappa\frac{X^2}{M_{\rm Pl}}H_uH_d ,
\eeq
where $X$ is an R-breaking spurion in the visible sector. Thus when $X$ acquires a VEV $\langle X\rangle\sim f_R$ this generates an effective $\mu$-term of order
\beq
 \mu   = \frac{ \kappa f_R^2}{M_{\rm Pl}} \gtrsim \kappa m_{3/2}~,
\eeq
where we use eq.~\eqref{eq:fR-bound} to obtain the latter inequality.
A complete model must specify the Higgs charges and the operator that generates $ W_{\rm KN}$, but it is compelling that an appropriate scale is already selected by the lower limit of the metastability bound on $f_R$.

In addition to forbidding the $\mu$-term, U(1)${}_R$ also prohibits Majorana gaugino masses. Accordingly, to obtain a phenomenologically viable R-symmetric SUSY extension of the Standard Model, it is common to introduce Dirac partners for each gaugino, see e.g.~\cite{Fayet:1975yi,Hall:1990hq,Fox:2002bu,Kribs:2007ac,Unwin:2012fj,Unwin:2024yqq}. Chiral superfields transforming in the adjoint are introduced for each gauge group such that the gauginos have Dirac masses arising from operators of the form
\beq
 \int d^2\theta 
 \frac{W_X^\alpha W^a_\alpha A^a}{M}~,
\eeq
where $A^a$ is the adjoint chiral multiplet. Since such operators require $D$-term spurions $W_X^\alpha$  the mixed $F/D$ structure of the present construction provides a natural fit. Parametrically,\footnote{This $D$-term relation breaksdown for the special case $r_*\approx3$ for which the D-term is effectively negligible.} since $D_X\sim m_{3/2}M_{\rm Pl}$, mediation at a scale $M\sim M_{\rm Pl}$ gives $m_D\sim m_{3/2}$. Thus,  the same type of uplift sector that helps realize the small-$f_R$ vacuum also provides the ingredients needed to make the R-symmetric visible sector phenomenologically viable. 

We conclude that, within the effective construction presented here, a QCD R-axion can be realized.  The price of obtaining a phenomenologically viable intermediate scale axion is a special structure for the superpotential and K\"ahler potential, a small parameter $\epsilon\ll1$, and the understanding that the model requires a consistent UV completion. While the model resolves the strong CP problem, by dynamically setting $\bar\theta$ to zero, it does not remove all small parameters since it requires a small parameter in the K\"ahler potential  $\epsilon\ll1$. However, this is a modest success since arguably small fundamental parameters are less concerning than an unexplained precision cancellation between two unrelated quantities, recall $ \bar\theta =  \theta+\arg\det M_q$. Moreover,  a small coefficient in the K\"ahler potential may plausibly arise from approximate symmetries or UV dynamics. For instance, similar type small couplings appear in supergravity models of inflation \cite{Brax:2005jv,Antusch:2009ef}. In these settings, shift symmetries of the K\"ahler potential are used to control the supergravity $\eta$ problem, with small parameters arising after symmetry breaking.

\section{Concluding remarks}
\label{S7}

This paper presents a new perspective on intermediate scale R-axions.  The usual expectation, following DFK \cite{Dine:2009sw}, is that in a nearly Minkowski vacuum with broken supersymmetry and broken R-symmetry, the R-axion decay constant is naturally of order $M_{\rm Pl}$.  We have presented a \textit{proof of principle} that this conclusion can be avoided locally in an effective construction if $D_TW=0$ such that the $T$ multiplet, which contains the R-Goldstone boson, is absent from the supersymmetry-breaking direction. Requirements on EFT validity and vacuum metastability reintroduce a bound on $f_R$, but this is found to be weaker than the naive DFK implication, specifically: $f_R\gtrsim\sqrt{m_{3/2}M_{\rm Pl}} $. Notably, for $m_{3/2}\sim1 {\rm TeV}$ the lower edge of this range lies near ${\rm few}\times 10^{11} {\rm GeV}$, which is typically the scale of concentrated interest for QCD axion models since it falls within the range for which the QCD axion can reproduce the dark matter relic abundance via the misalignment mechanism.

It should be noted that quantum gravity is not expected to preserve global symmetries \cite{Banks:2010zn,Harlow:2018tng}, thus Planck-suppressed R-violating operators may still spoil the PQ mechanism unless the global R-symmetry is protected \cite{Georgi:1981pu,Kamionkowski:1992mf,Holman:1992us,Barr:1992qq,Dobrescu:1996jp}. Discrete or continuous gauge symmetries in the UV completion can potentially forbid these problematic $M_{\rm Pl}$-induced operators. A novel prospect in this model is that the global R-symmetry could be a remnant of a UV gauged R-symmetry once UV is completed into supergravity (although the local U(1)$_R$ and remnant global symmetries should be distinct to avoid the R-axion of the latter being eaten). It is also worth noting that it has been suggested that $M_{\rm Pl}$-induced operators could be dressed by exponentially small coefficients, in which case the quality problem is largely moot \cite{Kallosh:1995hi}.

We highlight that a limitation of the uplift sector involves an effective FI parameter, which requires UV completion \cite{Dvali:1996rj,Arkani-Hamed:1998ufq}. In particular, the high energy theory must generate the required uplift without gauging away the R-axion, without inducing a large explicit axion potential, and without spoiling the condition $D_TW=0$ at the small-$f_R$ point.  Furthermore, we recall that our construction generically required the $D$-term to be associated with an ultraweak gauge coupling, $ g\sim m_{3/2}/M_{\rm Pl}$. These features are acceptable within the EFT, but they may be problematic for finding a consistent UV completion into quantum gravity \cite{Vafa:2005ui}. Such an ultraweak gauge coupling typically runs into weak-gravity-type constraints \cite{Arkani-Hamed:2006emk,Heidenreich:2016aqi,Craig:2019zkf,Daus:2020vtf}. 
It is interesting to note that Daus \textit{et al.} \cite{Daus:2020vtf} showed that approximate global symmetries descending from a UV gauge symmetry can be quantitatively constrained by swampland arguments.  In particular, for this class of gauge-derived approximate global symmetries, instanton-induced breaking effects should be expected with suppressions no smaller than
\beq
\exp\left[-M_{\rm Pl}^2/\Lambda^2\right]\sim\exp\left[-M_{\rm Pl}^4/f_R^4\right],
\eeq
where $\Lambda$ is the cutoff of the four-dimensional EFT, and in the latter expression we have substituted the relevant cutoff for the present model $\Lambda\sim f_R^2/M_{\rm Pl}$, as identified in eq.~\eqref{La}.  Thus, the low cutoff associated with the small-$f_R$ EFT permits the dangerous R/PQ-violating effects to be exponentially smaller, provided the low-energy global R/PQ symmetry descends from a UV gauge structure.

This work has presented a \textit{proof-of-principle} construction of an intermediate scale R-axion, at least at the EFT level.   The effective theory constructed here therefore provides a counterexample to the strongest folk version of the claim that $f_R\gtrsim M_{\rm Pl}$. At the same time, the construction makes clear that the Planckian expectation is difficult to evade: the intermediate scale regime is only metastable and depends on UV ingredients that are not automatic.  In closing, we note that intermediate scale R-axions may have applications beyond the QCD axion.  Previous work has explored R-axions in early-universe cosmology, including non-Gaussianity \cite{Nakayama:2009cr}, R-symmetric axionic moduli in supergravity models of natural inflation \cite{Harigaya:2014ola}, and the classic cosmological role of R-axions in dynamical supersymmetry breaking \cite{Bagger:1994hh,Nelson:1993nf}.  These examples suggest that scenarios with intermediate scale R-axions may have broader implications beyond novel implementations of the PQ mechanism emphasized here.

 \vspace{2mm}\noindent {\bf Acknowledgments.}~We thank Keisuke Harigaya, Arthur Hebecker, and Chang Sub Shin for useful interactions. JU is supported by NSF grant PHY-2209998.


\appendix 
\section{Appendices}
\subsection{Failure of the $D$-term uplift model}
\label{SA}

In the main text, we used a mixed $F/D$ uplift to obtain a local small-$f_R$ Minkowski vacuum.  In this appendix we show why the simpler constant $D$-term uplift construction is not sufficient.   The basic idea is reminiscent of the Dvali-Pomarol $D$-term-type model \cite{Dvali:1997sf}. We first construct a sector with $ D_TW=0$ and $W\neq0$, so that the modulus sector is supersymmetric even though the gravitino mass is nonzero.  

As in the main text, we take a modulus $T=s+i\tau$, whose imaginary part is the would-be R-axion, with the superpotential (identical to eq.~(\ref{W}))
\beq
 W(T)=W_*e^{2T}.
\eeq
In contrast to the main text there is no additional matter uplift sector (i.e.~no $\phi$).
We take K\"ahler potential to be
\beq
 K^{(0)}= M_{\rm Pl}^2
 \left[
  -2(T+T^\dagger)+ \frac{\epsilon^2}{2}(T+T^\dagger)^2+  \frac{c}{4}(T+T^\dagger)^4
 \right],
\eeq
with $ \epsilon\ll1$, and $c>0$. Writing  $T+T^\dagger=2s$, the relevant derivatives are
\beq
 \frac{K_T}{M_{\rm Pl}^2}= -2+2\epsilon^2s+8cs^3,
 \qquad
 \frac{K_{T\bar T}}{M_{\rm Pl}^2}= \epsilon^2+12cs^2 .
 \label{eq:ApDev}
\eeq
At the candidate critical point $s=0$
\beq
 K_T=-2M_{\rm Pl}^2,
 \qquad
 K_{T\bar T}=\epsilon^2M_{\rm Pl}^2 .
\eeq
Then the K\"ahler-covariant derivative of the superpotential is
\beq
 D_TW \big|_{s=0}
 =\left[
 \partial_TW+ \frac{K_T}{M_{\rm Pl}^2}W \right]_{s=0} = 0.
\eeq
Thus, similar to the main text, the $T$ sector is supersymmetric at $s=0$, while $W\neq0$.
Also, as in eq.~\eqref{fR}, the decay constant follows from the kinetic term, which at $s=0$ leads to
$  f_R(0)  =  \sqrt{2}\epsilon M_{\rm Pl}$ and for $\epsilon\ll1$, this is parametrically below $M_{\rm Pl}$.

We now check whether a constant uplift can lead to a stable Minkowski vacuum.  Adding a constant positive contribution to the scalar potential
\beq
 V_{\rm up}= \frac12D_X^2 ,
 \label{eq:ApTune}
\eeq
where $D_X$ is a constant, unlike the field-dependent $\mathcal D_X$ of the main text, see eq.~(\ref{eq:D}).
For the vacuum energy to be tuned to zero at $s=0$ requires
\beq
 \frac12D_X^2= 3m_{3/2}^2M_{\rm Pl}^2 .
\eeq
Thus, this construction successfully realizes  $D_TW=0$, and $ V_F+V_{\rm up}=0 $ (analogous to the model in the main text), however, unlike the mixed $F/D$ uplift, with the constant $D$-term uplift alone, this critical point is not stable. To see the instability, we compute the saxion potential away from the origin.
Using eq.~\eqref{eq:ApDev}, we have for general $s$
\beq
 D_TW=2s(\epsilon^2+4cs^2)W ,
\eeq
and the $F$-term potential along the saxion direction is
\beq
 V_F(s)=m_{3/2}^2M_{\rm Pl}^2
 \exp\left[
  2\epsilon^2s^2+ 4cs^4
 \right]
 \left[
  \frac{4s^2(\epsilon^2+4cs^2)^2}
  {\epsilon^2+12cs^2} -3
 \right],
\eeq
where $m_{3/2}$ denotes the gravitino mass evaluated at $s=0$ for which
\beq
 m_{3/2}=\frac{|W_*|}{M_{\rm Pl}^2}.
\eeq
Note that the  $F$-term potential matches eq.~\eqref{eq:VFx} with $r$ set to zero.

Since the potential is even in $s$, it follows that $s=0$ is a critical point.  Expanding around the
origin gives
\beq
 V_F(s)  \simeq m_{3/2}^2M_{\rm Pl}^2 \left[ -3-2\epsilon^2s^2+  {\cal O}(s^4) \right].
\eeq
Identical to eq.~\eqref{hat}, the canonically normalized saxion is
\beq
 \sigma =\sqrt{2}\epsilon M_{\rm Pl}s .
\eeq
Thus the quadratic term in the potential is
\beq
 V_F  = V_F(0) - m_{3/2}^2\sigma^2 + \cdots .
\eeq
Comparing with the canonical form $ V=V_0 + \frac12m_\sigma^2\sigma^2 +\cdots $ we find
\beq
 m_\sigma^2 =-2m_{3/2}^2 .
\eeq
Since the uplift in eq.~\eqref{eq:ApTune}  adds only a constant positive term to the potential, $ V(s)= V_F(s)+V_{\rm up}$, whilst it can tune the value of the vacuum energy at $s=0$, it cannot impact the derivative of the potential with respect to $s$, i.e.
\beq
 \frac{\partial^2 V}{\partial s^2}= \frac{\partial^2 V_F}{\partial s^2}.
\eeq
As such the saxion mass is unchanged by the uplift and remains tachyonic.  Thus we conclude that the saxion is unavoidably tachyonic, a constant uplift cannot produce a stable small-$f_R$ Minkowski vacuum, and this (simpler) model fails. Contrast this with eq.~\eqref{ms} in which the $r$ dependence provides an additional degree of freedom $m_\sigma^2= 2(r_\star-1)m_{3/2}^2$, thus resolving this issue.

\subsection{Stability of the uplifted stationary point at $s=0$}
\label{ApA}

Completing the analysis of Section~\ref{S3}, we verify that the uplifted stationary point at $s=0$ in the mixed $F/D$ uplift model is a local minimum.  The saxion direction was analyzed in the main text, so it remains to check the radial uplift direction. Near the stationary point we parameterise
\beq
\phi=
\left(\sqrt{r_\star}M_{\rm Pl}+\frac{\rho}{\sqrt2}\right)e^{i\theta_\phi}.
\eeq
Expanding $r$ at $s=0$ gives
\beq
r=\frac{|\phi|^2}{M_{\rm Pl}^2}
=r_\star+\frac{\sqrt{2r_\star}}{M_{\rm Pl}}\rho+\frac{\rho^2}{2M_{\rm Pl}^2}+\cdots .
\eeq
Since $\phi$ has the canonical K\"ahler potential of eq.~\eqref{K}, the radial field $\rho$ is canonically normalized.  Using the expansion above, and evaluating at the stationary point, the physical $\rho$ mass is
\beq
m_{\rho,\star}^2=\frac{2r_\star}{M_{\rm Pl}^2}\left.\frac{\partial^2 V_{s=0}}{\partial r^2}\right|_{r=r_\star}.
\eeq

At $s=0$, the scalar potential is given by eq.~\eqref{Vs0}, and differentiating twice with respect to $r$ gives
\beq
\frac{\partial^2 V|_{s=0}}{\partial r^2}=
(r-1)e^r\frac{|W_*|^2}{M_{\rm Pl}^2}
+g^2M_{\rm Pl}^4 .
\eeq
Thus, at the stationary point,
\beq
m_{\rho,\star}^2=2r_\star\left[(r_\star-1)m_{3/2}^2+g^2M_{\rm Pl}^2\right].
\eeq
The radial uplift direction is therefore locally stable for $r_\star>1$.

There is also no quadratic mixing between the saxion and the radial uplift mode.  Since the potential is an even function of $s$ near $s=0$, one has
\beq
\left.\frac{\partial^2 V}{\partial s\partial r}\right|_{s=0,r=r_\star}=0 .
\eeq
The Hessian block of the radial modes therefore diagonalizes at $s=0$.  The Hessian is positive semi-definite, with two zero modes corresponding to the physical R-axion and the U(1)$_X$ Goldstone mode which are subsequently lifted or eaten.  Thus the uplifted stationary point is a local minimum, provided
\beq
1<r_\star<3 .
\eeq
The upper bound follows from requiring the Minkowski condition in eq.~\eqref{eq:Mink} to have a real positive solution, while the lower bound comes from stabilization of the saxion.

\subsection{Trajectory away from the local minimum}
\label{ApB}

In Section~\ref{S3} we outlined a procedure for studying trajectories in field space away from the local minimum at $s=0$. Specifically, for each value of the saxion we minimize the potential with respect to the uplift variable $r$ to obtain an effective trajectory in field space
\beq
 V_{\rm eff}(s)=  \min_r V(s,r).
 \label{A1}
\eeq
The field $\phi$ is then allowed to adjust so as to minimize the potential at each value of $s$, the vacuum energy is tuned to vanish at the local minimum $s=0$. In this Appendix, we show in the generic case (we discuss the special case of $r_*\approx3$ in Appendix \ref{ApD}) that there is a deeper region of the potential away from $s=0$, towards $s_0\sim1$. Additionally, for larger $s$ the minimum in the $r$ direction shifts toward smaller $r$, i.e.~smaller values of $|\phi|$.

We denote by $r_{\rm min}(s)$ the value of $r$ obtained by minimizing the potential at fixed $s$:
\beq
\left.\frac{\partial V}{\partial r} \right|_{r=r_{\rm min}(s)}=0 .
\label{der}
\eeq
Since the potential is even in $s$, the minimizing trajectory can be expanded as
\beq
r_{\rm min}(s)=r_\star+C s^2+{\cal O}(s^4),
\eeq
where $C$ determines the leading impact of the uplift field to a saxion displacement.  Recall the $F$-term potential of eq.~(\ref{VF}) which we now express as
\beq
V_F\simeq m_{3/2}^2(s,r)M_{\rm Pl}^2\left[D(s)+r-3\right],
\eeq
where
\beq
D(s)=\frac{4s^2(\epsilon^2+4cs^2)^2}{\epsilon^2+12cs^2}\simeq 4\epsilon^2s^2+{\cal O}(s^4),
\eeq
and the field-dependent gravitino mass has the form
\beq
m_{3/2}^2(s,r)=e^{r+2\epsilon^2s^2+4cs^4}\frac{|W_*|^2}{M_{\rm Pl}^4}\simeq m_{3/2}^2(0,r) \left[1+2\epsilon^2s^2+{\cal O}(s^4)\right].
\eeq
Differentiating we obtain 
\beq
\frac{\partial V}{\partial r}=m_{3/2}^2(s,r)M_{\rm Pl}^2
\left[D(s)+r-2\right]+g^2M_{\rm Pl}^2\left(rM_{\rm Pl}^2+\xi\right).
\eeq
Substituting the $r_{\rm min}$ expansion into the stationarity condition and expanding to order $s^2$, the zeroth-order
piece vanishes by stationarity at $s=0$.  Thus to satisfy the stationarity condition at order $s^2$ we require
\beq
2\epsilon^2 r_\star m_{3/2}^2M_{\rm Pl}^2 +C\left[(r_\star-1)m_{3/2}^2M_{\rm Pl}^2+g^2M_{\rm Pl}^4\right]=0.
\label{00}
\eeq
Solving eq.~(\ref{00}) for $C$ gives
\beq
C=-\frac{2\epsilon^2 r_\star m_{3/2}^2M_{\rm Pl}^2}{(r_\star-1)m_{3/2}^2M_{\rm Pl}^2+g^2M_{\rm Pl}^4}.
\label{eq:drmin}
\eeq
For $r_\star>1$, as required for saxion stability, the denominator is positive and  $C<0$. 
Moreover, since $C$ is the coefficient of the leading $s^2$ shift in $r_{\rm min}(s)$, this implies
\beq
C =\left.\frac{dr_{\rm min}}{d(s^2)} \right|_{s=0}.
\eeq
Thus the minimizing trajectory moves toward smaller $r$ as the saxion is displaced from the small-$f_R$ point.  Since $r=|\phi|^2/M_{\rm Pl}^2$, this means that the uplift field $\phi$ moves toward a smaller expectation value, reducing the uplift-sector $F$-term contribution.

The above shows the initial direction in which the uplift field $\phi$ moves as the saxion is displaced. To determine whether this direction connects to a lower-energy region, one must examine the potential away from the immediate neighborhood of $s=0$. In this regime, the terms in the K\"ahler potential that were negligible near the origin become important.
  In particular, since
\beq
 K_{T\bar T}=M_{\rm Pl}^2
 \left(  \epsilon^2+12cs^2 \right),
\eeq
 the transition from the small $f_R$ regime to the generic Planckian-$f_R$ regime occurs when $12cs^2 \sim 1$, and for  $c\sim1$, this corresponds to $ s\sim1$.
As discussed in Section~\ref{S5},  away from $s=0$ the special cancellation in the K\"ahler-covariant derivative no longer occurs, and for $s\neq0$ one has $D_TW\neq0$.  Moreover, for $s\sim 1$ with  $c\sim1$, the R-axion scale also becomes Planckian
\beq
 f_R(s)= \sqrt{2K_{T\bar T}}= \sqrt{2}M_{\rm Pl}
 \sqrt{\epsilon^2+12cs^2}
 \sim M_{\rm Pl}.
\eeq
  Thus the lower-energy trajectory, after leaving the metastable small-$f_R$ region, naturally approaches the order-one saxion regime.

It remains to ascertain whether the $s=0$ or $s\sim1$ vacuum is deeper.
Notably, since $  V_{\rm eff}(s)\leq V(s,0)$, if $V(s,0)<0$ for some value of $s$, then the tuned Minkowski point at $s=0$ cannot be the global minimum.
Let $r_\star$ denote the location of the metastable Minkowski point. Using the Minkowski and stationarity conditions at $s=0$, one has (from eq.~(\ref{g}))
\beq
 \frac{g^2M_{\rm Pl}^2}{m_{3/2}^2}
 = \frac{(r_\star-2)^2}{2(3-r_\star)}
 \label{eq:g_relation_for_depth}
\eeq
and (from eq.~(\ref{xi}))
\beq
 \frac{\xi}{M_{\rm Pl}^2}
 = -r_\star- \frac{(r_\star-2)m_{3/2}^2}{g^2M_{\rm Pl}^2}.
 \label{eq:xi_relation_for_depth}
\eeq
Therefore, along a test trajectory with $r$ fixed to $r=0$, the potential can be written as
\beq
 \frac{V(s,0)}{m_{3/2}^2M_{\rm Pl}^2}
 = e^{-r_\star}
 \exp\left[   2\epsilon^2s^2+4cs^4   \right]
 \left[  \frac{4s^2(\epsilon^2+4cs^2)^2}{\epsilon^2+12cs^2}
  - 3 \right] +   \frac{g^2M_{\rm Pl}^2}{2m_{3/2}^2}
 \left( \frac{\xi}{M_{\rm Pl}^2}\right)^2 .
 \label{eq:trial_potential_r_zero}
\eeq
The second term is fixed entirely by the uplift conditions. 
This term diverges as $r_\star\rightarrow3$, indicating the large $D$-term cost of the $r=0$ trajectory in this limit. This is consistent with
the special case discussed in Appendix~\ref{ApD}.
Furthermore, for $\epsilon\ll1$, the first term in eq.~\eqref{eq:trial_potential_r_zero}
simplifies.  Writing $ y\equiv4cs^4$, gives
\beq
 \frac{V(s,0)}{m_{3/2}^2M_{\rm Pl}^2}
 \simeq
 e^{-r_\star}e^y
 \left(  \frac{4y}{3}-3  \right) +  \frac{(r_\star^2-4r_\star+6)^2}{4(3-r_\star)} ,
 \label{eq:trial_potential_y}
\eeq
where we use \eqref{eq:g_relation_for_depth} and \eqref{eq:xi_relation_for_depth} to eliminate $g$ and $\xi$.
The first term, which contains the only $y$ dependence, is minimized at $   y=5/4$, corresponding to
\beq
 s_0=\left(\frac{5}{16c} \right)^{1/4}.
 \label{eq:s0_order_one}
\eeq
For $c\sim1$, this is an order-one displacement in the saxion variable $s$, and at this point
\beq
 \frac{V(s_0,0)}{m_{3/2}^2M_{\rm Pl}^2}
 \simeq
 \frac{(r_\star^2-4r_\star+6)^2}{4(3-r_\star)}-\frac{4}{3}
 e^{5/4-r_\star}.
 \label{eq:trial_depth_condition}
\eeq
Whenever the right-hand side is negative, the point $(s_0,r=0)$ lies below the metastable Minkowski vacuum.  
Since $V_{\rm eff}(s_0)\leq V(s_0,0)$, this is a sufficient condition for the existence of a deeper region of the effective potential. Whenever this condition is satisfied, the small-$f_R$ point is therefore a metastable local minimum rather than the global minimum of the effective potential. This conclusion generically applies except for the special case in which the uplift sector stabilises at $r_\star\simeq3$, as we discuss next.

\subsection{The $r_\star\approx 3$ limit of the uplift sector}
\label{ApD}

We highlighted above that $r_\star=3-\delta$ avoided an ultraweak gauge coupling (for $\delta>0$).
The character of the uplift also changes in this limit.  The $D$-term contribution to the positive vacuum energy is
\beq
 V_D(r_\star) = (3-r_\star)m_{3/2}^2M_{\rm Pl}^2
 = \delta m_{3/2}^2M_{\rm Pl}^2 .
\eeq
Thus, as $\delta\rightarrow0$, the vacuum energy is cancelled almost entirely by the $\phi$ $F$-term contribution, while the $D$-term contribution tends to zero. 

Accordingly, some aspects of the analysis are altered. The local stability analysis of Section \ref{S3} and Appendix~\ref{ApA} is largely unchanged, however the trajectory away from the local minimum is modified.  Near $s=0$, the shift of the $r$ minimum is given by eq.~\eqref{eq:drmin}, using eq.~\eqref{g} and taking $r_\star=3-\delta$ this can be re-expressed as
\beq  \left.   \frac{dr_{\rm min}}{d(s^2)}  \right|_{s=0} \simeq -\frac{2\epsilon^2(3-\delta)}{1/(2\delta)} \simeq -12\epsilon^2\delta .
\eeq
This is because the denominator in this case is dominated by the $g^2M_{\rm Pl}^4$ term. The change in the minimum in the $r$ direction as the saxion is displaced is suppressed by $\delta\ll1$ compared to the generic case with $1<r_\star<3$. The $r\approx0$ trial trajectory used in Appendix~\ref{ApB} is no longer useful, as its $D$-term cost grows parametrically as $1/\delta$.
Thus the lower-energy path which favoured larger $s$ in the generic case is absent for $r_\star\approx3$. This suggests that $s=0$ may be the global minimum (although we do not prove that here, in particular, the global minimum may be sensitive to higher-order K\"ahler corrections to the uplift sector since $r$ is large).
Interestingly, if $s=0$ is a global minimum, this would imply that it is an absolutely stable vacuum and the metastability bounds of Section \ref{S5} do not apply to this special case.

Finally, we examine the EFT constraint of eq.~(\ref{EFT2}), for the case of $r_\star=3-\delta$ this is
\beq
 f_R\gtrsim \sqrt{m_{3/2}M_{\rm Pl}}.
\eeq
Notably, up to $\mathcal{O}(1)$ factors, this coincides with the lower bound of eq.~\eqref{eq:fR-bound} coming from the vacuum metastability limits relevant for $r_\star\not\approx3$.


\end{document}